\documentclass[aps,twocolumn,groupedaddress,floatfix, longbibliography]{revtex4-1}
\usepackage{bm}
\usepackage{amsmath}
\usepackage{amsfonts}
\usepackage[colorlinks=true, allcolors=blue]{hyperref}

\newcommand{\bx}{\bm{x}}
\newcommand{\bq}{\bm{q}}
\newcommand{\bk}{\bm{k}}
\newcommand{\bg}{\bm{g}}
\newcommand{\bt}{\bm{t}}
\newcommand{\bn}{\bm{n}}
\newcommand{\bnu}{\bm{\nu}}
\newcommand{\br}{\bm{r}}
\newcommand{\bu}{\bm{u}}
\newcommand{\bQ}{\bm{Q}}

\begin{document}
\title{Thermal fluctuations in crystalline membranes with long-range dipole interactions}
\author{Achille Mauri}
\email[]{a.mauri@science.ru.nl}
\author{Mikhail I. Katsnelson}
\affiliation{Radboud University, Institute for Molecules and Materials, NL-6525AJ Nijmegen, The Netherlands}
\date{\today}

\begin{abstract}
We study the effects of long-range electrostatic interactions on the thermal fluctuations of free-standing crystalline membranes exhibiting spontaneous electric polarization directed at each point along the local normal to the surface. We show that the leading effect of dipole-dipole interactions in the long-wavelength limit consists in renormalizations of the bending rigidity and  the elastic coefficients. A completely different result was obtained in the case of scalar two-point interactions decaying as $\left|\bm{R}\right|^{-3}$, where $\left|\bm{R}\right|$ is the distance. In the latter case, which was addressed in previous theoretical research, the energy of long-wavelength bending fluctuations is controlled  by power-law interactions and it scales with the wavevector $\bm{q}$ as $\left|\bm{q}\right|^{3}$, leading to a modified large-distance behaviour of correlation functions.  By contrast, in the case of dipole interactions, the $\left|\bm{q}\right|^{3}$ dependence of the bending energy vanishes. Non-local terms generated by the expansion of the electrostatic energy are suppressed in the limit of small wavevectors. This suggests that the universal scaling behaviour of elastic membranes holds even in presence of dipole interactions. At the same time, the shift of the Lam\'e coefficients and the bending rigidity induced by electrostatic interactions can be quantitatively important for two-dimensional materials with a permanent out-of-plane polarization.
\end{abstract}
\maketitle

\section{Introduction \label{introduction}}
The thermodynamic properties and the elastic response of two-dimensional (2D) crystalline membranes subject to small or vanishing external tension are crucially determined by the thermal fluctuations of their shape. The statistical mechanics of fluctuating elastic membranes has been the subject of extensive research effort~\cite{nelson_statistical_1989, katsnelson_graphene:_2012, bowick_statistical_2001, nelson_fluctuations_1987, aronovitz_fluctuations_1988, guitter_thermodynamical_1989, le_doussal_self-consistent_1992, le_doussal_anomalous_2018, gazit_structure_2009, kownacki_crumpling_2009, kosmrlj_mechanical_2013, kosmrlj_response_2016, burmistrov_quantum_2016, gornyi_rippling_2015, gornyi_anomalous_2017, katsnelson_graphene_2013, los_mechanics_2017, los_scaling_2016, burmistrov_stress-controlled_2018}. The problem is particularly complex because of low dimensionality and the softness of the dispersion relation of bending modes, which leads to strong fluctuations and the breakdown of the harmonic approximation. The energy and the nature of fluctuations depend essentially on the degree of internal order of the membrane~\cite{nelson_statistical_1989}. In the case of fluid membranes, the shear modulus vanishes. It was demonstrated that bending fluctuations then lead to a crumpled state at any finite temperature~\cite{nelson_statistical_1989}. By contrast, membranes with an internal crystalline order and, therefore, a finite shear modulus present a low-temperature flat phase, characterized by a planar average configuration. The stability of  the flat phase long-range order to thermal bending fluctuations is made possible by the anharmonic coupling of out-of-plane and in-plane modes~\cite{nelson_fluctuations_1987}. The statistical mechanical properties of the flat phase are controlled by long-wavelength fluctuations of out-of-plane (bending) and in-plane (stretching) modes. In absence of long-range interactions and anisotropies in the elastic coefficients, the most general effective Hamiltonian for slowly varying deformations is~\cite{nelson_statistical_1989, nelson_fluctuations_1987, aronovitz_fluctuations_1988, guitter_thermodynamical_1989, le_doussal_self-consistent_1992}
\begin{equation}
\label{eq:elasticity_hamiltonian}
H_{\text{el}} = \int {\rm d}^{2}x \left[\frac{1}{2}\kappa \left(\nabla^{2}h\right)^{2} + \frac{\lambda}{2}\tilde{u}_{\alpha \alpha}^{2} + \mu \,\tilde{u}_{\alpha \beta} \tilde{u}_{\alpha \beta}\right]~, 
\end{equation}
where $h(\bx)$ and $\bu(\bx)$ are out-of-plane  and in-plane displacements and
\begin{equation}
\label{eq:u_tilde}
\tilde{u}_{\alpha \beta} = \frac{1}{2}(\partial_{\alpha}u_{\beta} + \partial_{\beta} u_{\alpha} + \partial_{\alpha} h \partial_{\beta} h )
\end{equation}
is the strain tensor. $\lambda$ and $\mu$ are Lam\'e coefficients and $\kappa$ is the bending rigidity. Terms including higher powers of $\partial_{\alpha} h$ or $\partial_{\alpha} u_{\beta} $ or higher-order gradients can be neglected, because they are irrelevant at long wavelengths~\cite{nelson_fluctuations_1987, aronovitz_fluctuations_1988, guitter_thermodynamical_1989, notequantum}. The field $ h(\bx)$ plays the role of the Goldstone mode related to broken rotational symmetry~\cite{guitter_thermodynamical_1989, le_doussal_anomalous_2018, bowick_statistical_2001} and it exhibits a particularly soft dispersion relation.
In the harmonic approximation, the energy of bending fluctuations scales as $\kappa q^{4}$, where $q$ is the wavevector. The anharmonic coupling between in-plane and out-of-plane modes constitutes a strongly relevant perturbation to the Gaussian fixed point, making the statistical mechanics very non-trivial~\cite{nelson_statistical_1989}. Correlation functions of $h$ and $\bu$ fields can be calculated perturbatively in the anharmonic coupling strength for wavevectors larger than the 'Ginzburg scale' $q_{*} \approx \sqrt{3Y/ 16\pi \kappa^{2}}$, with  $Y = 4\mu (\lambda + \mu) /(\lambda + 2\mu)$. For $q \ll q_{*}$, instead,  perturbation theory breaks down~\cite{nelson_fluctuations_1987, katsnelson_graphene:_2012, katsnelson_graphene_2013}. The long-wavelength behaviour of correlation functions is determined by a non-trivial fixed point Hamiltonian, characterized by anomalous scaling exponents~\cite{nelson_statistical_1989,  bowick_statistical_2001, nelson_fluctuations_1987, aronovitz_fluctuations_1988, guitter_thermodynamical_1989, kownacki_crumpling_2009}. In the small-$q$ regime, the bending rigidity diverges as $q^{-\eta}$ and the elastic Lam\'e coefficients both vanish as $q^{2-2\eta}$, where $\eta$ is an universal critical exponent. The determination of $\eta$ has been addressed through several field-theoretic methods~\cite{nelson_statistical_1989, aronovitz_fluctuations_1988, guitter_thermodynamical_1989, le_doussal_self-consistent_1992, le_doussal_anomalous_2018, gazit_structure_2009, kownacki_crumpling_2009}. The self-consistent screening approximation~\cite{le_doussal_self-consistent_1992, le_doussal_anomalous_2018} and the functional renormalization group~\cite{kownacki_crumpling_2009} lead to $\eta \simeq 0.821$ and $\eta \simeq 0.849$, respectively.

The anomalous scaling of the elastic and bending coefficients has crucial consequences in the physical properties of crystalline membranes. The very stability of the flat phase and the existence of long-range order in the orientation of the normal to the membrane relies on the power-law stiffening of the bending rigidity and would be impossible if anharmonic effects were neglected~\cite{nelson_fluctuations_1987}. In addition, the renormalization of elastic coefficients determines the macroscopic mechanical properties of crystalline membranes in an essential way, leading to anomalous strain and size dependent elastic response~\cite{los_mechanics_2017, los_scaling_2016, kosmrlj_mechanical_2013, kosmrlj_response_2016, katsnelson_graphene_2013, gornyi_anomalous_2017, burmistrov_stress-controlled_2018}.

The statistical mechanics of anharmonic thermal fluctuations of 2D crystals in the flat phase has fundamental importance in the theory of thermodynamic and mechanical properties of free-standing graphene or suspended samples of other atomically-thin materials~\cite{katsnelson_graphene:_2012, katsnelson_graphene_2013, los_mechanics_2017}. The Ginzburg wavelength $\lambda_{*} = 2\pi/q_{*}$ for these materials at room temperature is of the order of 10 \AA, which makes the anomalous scaling behaviour important already at short length scales~\cite{le_doussal_anomalous_2018}. The elasticity of graphene is currently under extensive investigation, both theoretically~\cite{los_scaling_2016, los_mechanics_2017, kosmrlj_response_2016} and experimentally~\cite{nicholl_effect_2015, lopez-polin_influence_2017}.

It has been understood long time ago that the scaling behaviour of crystalline membranes can be modified in presence of long-range interactions which decay sufficiently slowly with the distance. The effects of power-law interactions of the form
\begin{equation*}
 V(\bm{R}) = A/\left|\bm{R}\right|^{\sigma}
\end{equation*}
on $D$-dimensional membranes fluctuating in a $d$-dimensional embedding space were studied in Refs.~\cite{toner_elastic_1989, kantor_statistical_1989, guitter_tethered_1992, mori_tethered_1993}. In Ref.~\cite{toner_elastic_1989}, in particular, the author focused on effects of long-range interactions in the flat phase and addressed the scaling behaviour for generic $D$, $d$ and $\sigma$.
It was shown that, for small fluctuations around the flat configuration, the dominant effect of long-range interactions is to modify the long-wavelength behaviour of the quadratic part of the bending energy, which acquires a wavevector dependence $ \left|\bm{q}\right|^{\sigma + 2 - D} $.  When $\sigma + 2 - D < 4-\eta$, where $\eta$ is the short-range critical exponent, long-range interactions dominate the long-wavelength behaviour  and the universal anomalous elasticity predicted by conventional membrane theory does not hold~\cite{toner_elastic_1989}. If $\sigma$ is decreased further, $\sigma < \frac{3}{2}D$, the bending stiffness becomes larger and the exponents approach the Gaussian ones. The case of scalar interactions decaying as $\left|\bm{R}\right|^{-3}$ in two dimensions, lies precisely at the boundary of the domain of validity of the Gaussian approximation. The renormalized in-plane elastic coefficients $\lambda (q)$ and $\mu(q)$ then vanish as $1\big/\ln\left(1/q\right)$~\cite{toner_elastic_1989}. 

The fact that long-range interactions can crucially affect the behaviour of systems with massless modes is very general. In two-dimensional spin systems, for example, dipole-dipole interactions lead to the breaking of the conditions of the Mermin-Wagner theorem and have dramatic effects on thermodynamic properties~\cite{grechnev_thermodynamics_2005}. Long-range dipolar interactions also play a crucial role in spin systems at the ferromagnetic phase transitions, leading to a modified critical behaviour~\cite{aharony_critical_1973, fisher_critical_1972-1}.

Electrostatic interactions also play an important role in charged fluid membranes immersed in electrolyte solutions. Vast research effort was devoted to the quantitative prediction of the renormalization of bending moduli induced by electrostatic interactions~\cite{daicic_bending_1996, fogden_bending_1997, fogden_electrostatic_1996, kumaran_effect_2001, loubet_electromechanics_2013, netz_buckling_2001, shojaei_effects_2016}.

Several theoretical studies also addressed the curvature moduli of membranes subject to van der Waals interactions~\cite{netz_buckling_2001, shojaei_effects_2016, dean_renormalization_2006} and the elasticity of flexoelectric membranes~\cite{peliti_fluctuations_1989, liu_flexoelectricity_2013}.

In this paper, we discuss the effect of dipole-dipole interactions on the fluctuations of free-standing crystalline membranes. We focus, in particular, on the case of two-dimensional crystals with a permanent polarization of fixed magnitude oriented along the local normal, which we describe as arising from point dipole moments located at the lattice sites. We show that dipole-dipole interactions lead to a very different result from the one obtained for scalar interactions decreasing with distance as $\left|\bm{R}\right|^{-3}$.  In the quadratic part of the energy, the non-analytic wavevector dependence proportional to $\left|\bm{q}\right|^{\sigma + 2 -D} = \left|\bm{q}\right|^{3}$ of the inverse $h$ field propagator vanishes exactly. Expanding the electrostatic energy generates, in the leading order in the long-wavelength limit, to a local functional of the strain tensor. Non-local terms, instead, constitute small corrections in the limit of long wavelengths and they are not expected to modify the large-distance behaviour. Electrostatic interactions, on the other hand, lead to a shift of the Lam\'e coefficients $\lambda$ and $\mu$ and to the bending rigidity $\kappa$, which can be quantitatively important in realistic two-dimensional materials.

Our study is relevant for 2D materials which lack symmetry under inversion and under reflection with respect to the crystal plane and which can exhibit, therefore, a finite out-of-plane polarization. Important examples are the graphene derivatives C$_{2}$F (fluorinated on one side), C$_{2} $HF (fluorinated on one side and hydrogenated on the opposite side) and C$_{2}$LiF~\cite{klintenberg_theoretical_2010}. In these materials, polarization arises from the charge transfer between carbon and halogen and alkali atoms. Other examples of two-dimensional materials exhibiting spontaneous out-of-plane polarization are MoSSe monolayer and multilayers~\cite{riis-jensen_efficient_2018}. Recently, out-of-plane spontaneous polarization and ferroelectric phenomena were observed in bi- and trilayer WTe$_{2}$, which were identified as rare examples of 2D ferroelectric metals~\cite{fei_ferroelectric_2018}.

Thermal fluctuations of membranes with a spontaneous polarization directed along the normal to the surface were investigated in the previous work~\cite{peliti_fluctuations_1989}. However, in Ref.~\cite{peliti_fluctuations_1989}, the expansion of the electrostatic energy up to second order in $h(\bx)$ was calculated only approximately. This lead to the prediction of a $\left|\bq\right|^{3}$ wavevector dependence of the bending energy, which contrasts with our results. Fluctuations of nearly-planar dipole layers were also investigated in Ref.~\cite{miller_stability_1974}, which reports  the calculation of the electrostatic energy up to second order in the out-of-plane fluctuation amplitude and to order $q^{2}$ for a dipole layer at the interface between media with two different dielectric constants $\epsilon_{a}$ and $\epsilon_{b}$. In our work we determined, in the case $\epsilon_{a} = \epsilon_{b}$,  the fluctuation energy to higher orders in $\bm{q}$ and in the fluctuation amplitudes, which is crucial for addressing the statistical mechanical properties of the system.

A comparison between our calculations and the results in Refs.~\cite{daicic_bending_1996, fogden_electrostatic_1996, fogden_bending_1997, kumaran_effect_2001, loubet_electromechanics_2013, netz_buckling_2001, shojaei_effects_2016, dean_renormalization_2006} is not direct because of the differences of the systems under examination. In the case of charged membranes immersed in solutions with added salt, Coulomb interactions are Debye-screened and become effectively short-ranged. In this case, effects of electrostatic interactions can a priori be encoded in a renormalization of local elastic coefficients, in the limit of long wavelengths~\cite{daicic_bending_1996, fogden_electrostatic_1996, kumaran_effect_2001, loubet_electromechanics_2013}. On the other hand, electrostatic interactions are unscreened in the case of membranes immersed in solutions without added salt, in which the only mobile ions are counterions which ensure an overall charge neutrality~\cite{fogden_electrostatic_1996, fogden_bending_1997}. In Refs.~\cite{fogden_electrostatic_1996, fogden_bending_1997}, bending fluctuations of two charged  membranes with the same charge and with intervening counterions were addressed. It was found that, for in-phase fluctuation modes of the two layers, the wavevector dependence of the energy is regular at $\bm{q} \rightarrow 0$ despite the long-range nature of the interactions. This finding is similar to our results for the quadratic part of the energy of out-of-plane fluctuations in a membrane with dipole-dipole interactions, although the two systems and the methods to describe them are very different.

\section{Hamiltonian for crystalline membranes with long-range interactions}
Fluctuating configurations of a crystalline membrane in the flat phase are described by specifying the displacement  of all points from their position in the minimal energy configuration. In the rest minimum energy state, the membrane lies in a plane, which we can choose as the $xy$ plane. Displacements in the membrane plane are denoted as $\bu\left(\bx\right)$ and displacements in the orthogonal $z$ direction are denoted as $h\left(\bx\right)$.  The position of points in the three-dimensional embedding space is denoted as $\br(\bx) = \bx + \bu(\bx) + h(\bx) \bm{e}_{z} $. 
When the crystal lattice is sufficiently isotropic and interactions are short-ranged, long-wavelength fluctuations can be described through the coarse-grained Hamiltonian~\cite{nelson_statistical_1989, nelson_fluctuations_1987, aronovitz_fluctuations_1988, guitter_thermodynamical_1989, gornyi_rippling_2015}:
\begin{equation}
\label{eq:H_sr_complete}
\begin{split}
H_{\text{sr}} & = \int {\rm d}^{2}x \bigg[\frac{1}{2}\kappa \left(\nabla^{2}\br\right)^{2} \\
&+ \frac{\lambda}{2}u_{\alpha \alpha}^{2} + \mu \,u_{\alpha \beta} u_{\alpha \beta} + \tau u_{\alpha \alpha}\bigg]~,\\
\end{split}
\end{equation}
where $\kappa$ is the bending rigidity, $\lambda$ is the first Lam\'e coefficient and $\mu$ the shear modulus. The strain tensor $u_{\alpha \beta}$ is defined as:
\begin{equation}
\begin{split}
u_{\alpha \beta} & = \frac{1}{2}\left(\partial_{\alpha}\br \cdot \partial_{\beta}\br - \delta_{\alpha \beta}\right)\\ 
&= \frac{1}{2} \left(\partial_{\alpha} u_{\beta} + \partial_{\beta} u_{\alpha} + \partial_{\alpha}h\, \partial_{\beta}h+ \partial_{\alpha} u_{\gamma}\, \partial_{\beta} u_{\gamma}\right) ~.
\end{split}
\end{equation}
The linear term
\begin{equation}
\int {\rm d}^{2}x \, \tau\, u_{\alpha \alpha}\left(\bx\right)
\end{equation}
represents an elastic tension. If only the short-range forces described by Eq.~\eqref{eq:H_sr_complete} were present, $\tau$ should vanish because $\bu(\bx) = h(\bx) = 0$ was defined as the state of minimum energy. However, dipole-dipole interactions will be included. These tend to stretch the membrane. Therefore, a finite value of $\tau$ must be taken into account in order to ensure that the reference configuration will correspond to the state of minimum energy of the \emph{total} Hamiltonian. The appearance of a finite elastic tension will be important in the following. As shown in Section~\ref{sec:harmonic_approximation} (see Eq.~\eqref{eq:H_lr1_2_fourier}), electrostatic interactions lead in the bending energy to a negative, quadratic-in-q term, proportional to
\begin{equation*}
    -\sum_{\bq} q^{2}\left|h_{\bq}\right|^{2}~.
\end{equation*}
This term will be exactly cancelled by the positive tension $\tau$, as it was already noticed in Ref.~\cite{toner_elastic_1989}. The exact cancellation is ensured by rotational symmetry and is related to the 'Goldstone' nature of the out-of-plane mode $h(\bx)$.

Eq.~\eqref{eq:H_sr_complete} can be simplified by neglecting in-plane anharmonicity and the term $\left(\nabla^{2} \bu\right)^{2}$, which can be shown to be irrelevant for the long-wavelength behaviour~\cite{nelson_statistical_1989, nelson_fluctuations_1987, aronovitz_fluctuations_1988, guitter_thermodynamical_1989, gornyi_rippling_2015}. If these terms are ignored, $H_{\text{sr}}$ becomes:
\begin{equation}
\label{eq:H_sr}
\begin{split}
H_{\text{sr}} & = \int {\rm d}^{2}x \bigg[\frac{1}{2}\kappa \left(\nabla^{2}h \right)^{2} \\
&+ \frac{\lambda}{2}\tilde{u}_{\alpha \alpha}^{2} + \mu \,\tilde{u}_{\alpha \beta} \tilde{u}_{\alpha \beta} + \tau u_{\alpha \alpha}\bigg]~,\\
\end{split}
\end{equation}
where the approximate strain tensor $\tilde{u}_{\alpha \beta}$ is defined as in Eq.~\eqref{eq:u_tilde}.

In presence of dipolar interactions, the Hamiltonian for displacement fields is:
\begin{equation}
\label{eq:H_lr_continuum}
\begin{split}
& H  = H_{\text{sr}} + H_{\text{lr}}~, \\
& H_{\text{lr}}  = \frac{1}{2} P^{2} \int {\rm d}^{2}x \int {\rm d}^{2}x' \bigg [\frac{\bn(\bx) \cdot \bn(\bx')}{\left|\br(\bx) - \br(\bx')\right|^{3}} \\
&\;\;\; \;\;\; - \frac{3\left(\bn(\bx)\cdot \bnu\right)\left(\bn(\bx')\cdot \bnu\right)}{\left|\br(\bx) - \br(\bx')\right|^{3}}\bigg]~.
\end{split}
\end{equation}
Here $\bn(\bx)$ is the normal to the membrane surface at the point $\bx$ and $\bnu$ is the unit vector:
\begin{equation}
\bnu \equiv \frac{\br(\bx) - \br(\bx')}{\left|\br(\bx) - \br(\bx')\right|}~.
\end{equation}
$P$ is the two-dimensional polarization density, defined as the dipole moment per unit area. The electrostatic energy defined in Eq.~\eqref{eq:H_lr_continuum} is actually divergent when $\bx'$ approaches $\bx$ and a short-distance cutoff must be introduced. Motivated by the study of ordered crystals such as CF, C$_{2}$HF or C$_{2}$LiF we take care of the divergence of the energy by the following procedure. We consider the polarization to arise from finite dipole moments located at the sites of a perfect lattice. The energy becomes:
\begin{equation}
\label{eq:H_lr_lattice}
\begin{split}
H_{\text{lr}} & = \frac{1}{2} \sum_{i}\sum_{j\neq i} p_{i}p_j\bigg[\frac{\bn_{i} \cdot \bn_{j}}{\left|\br_{i}- \br_{j}\right|^{3}} \\
& - \frac{3\left(\bn_{i}\cdot \bnu\right)\left(\bn_{j}\cdot \bnu\right)}{\left|\br_{i} - \br_{j}\right|^{3}}\bigg]~.
\end{split}
\end{equation}The discrete indices $i$ and $j$ label the lattice sites at which dipole moments are located, $p_{i}$ denotes the magnitude of the dipole moment at the site $i$. $p_{i}$ is assumed to have the periodicity of the lattice, so that the dipole moment distribution is identical in all unit cells. For simplicity, despite the discretization of the problem, we still describe configurations through functions $\bu(\bx)$ and $h(\bx)$ defined on the continuum two-dimensional space. For consistency, the  functions $\bu(\bx)$ and $h(\bx)$ must be slowly varying, with finite Fourier components only for wavevectors well within the first Brillouin zone. In Eq.~\eqref{eq:H_lr_lattice}, we identify $\br_{i}$ and $\bn_{i}$ with $\br(\bx_{i})$ and $\bn(\bx_{i})$. This corresponds to picturing the membrane as a 2D continuum elastic medium in which dipole moments are embedded at discrete lattice positions~\cite{tamm_fundamentals_1979, *purcell_electricity_1965}. If the displacement fields vary slowly with respect to the atomic scale, the approximate character of the continuum description is inessential. It allows however a simpler definition of the unit vector $\bn$ normal to the surface and it avoids the necessity to derive a discretized version of the short-range elastic Hamiltonian in Eq.~\eqref{eq:H_sr}.

To study how electrostatic interactions perturb the properties of membranes which lie deep in the flat phase and far from the crumpling transition point, it is legitimate to expand Eq.~\eqref{eq:H_lr_lattice} in powers of the displacement fields $h(\bx)$, $\bu(\bx)$. According to the conventional statistical mechanics of membranes, if only the elastic energy, Eq.~\eqref{eq:H_sr}, is considered, the fluctuating field correlation functions scale at large distances according to:
\begin{equation}
\begin{split}
&\langle \left(h\left(\bx\right) - h\left(\bx'\right)\right)^{2}\rangle \approx \left|\bx - \bx'\right|^{2-\eta}\\
&\langle  \left(\bu\left(\bx\right) - \bu\left(\bx'\right)\right)^{2}\rangle \approx \left|\bx - \bx'\right|^{2-2\eta}~,
\end{split}
\end{equation}
with $\eta \simeq 0.8 $. Therefore, distances fluctuate much less than their average value. In addition, the mean square fluctuation of the normal around the average $z$-axis is finite and small if $T\ll  \kappa \approx T_{\text{cr}}$. This is the case for graphene and most two-dimensional materials at room temperature.

\subsection{Electrostatic energy: harmonic approximation}
\label{sec:harmonic_approximation}
Expanding Eq.~\eqref{eq:H_lr_lattice} up to second order in the fields gives the long-range contribution to the Gaussian fluctuation theory. It is convenient to rewrite the expression for the energy as:
\begin{equation}
\label{eq:H_lr_splitted}
\begin{split}
H_{\text{lr}} & = \frac{1}{2}\sum_{i}\sum_{j \neq i} p_{i} p_{j}\frac{1}{\left|\br_{i}-\br_{j}\right|^{3}} \\
& - \frac{1}{2}  \sum_{i}\sum_{j \neq i} p_{i}p_{j} \bigg[ \frac{\frac{1}{2}\left(\bn_{i}-\bn_{j}\right)^{2} + 3\left(\bn_{i}\cdot \bnu\right)\left(\bn_{j}\cdot \bnu\right)}{\left|\br_{i}-\br_{j}\right|^{3}}\bigg] \\
& \equiv H_{\text{lr},1} + H_{\text{lr},2}~.
\end{split}
\end{equation}The second line in Eq.~\eqref{eq:H_lr_splitted}, denoted $H_{\text{lr},2 }$, is now regular when $\bx_{i}$ approaches $\bx_{j}$ and the lattice sums can be safely transformed into integrals. To first order in the displacement fields, the normal $\bn(\bx)$ can be written as:
\begin{equation}
\label{eq:normal_1o}
\bn(\bx) = \bm{e}_{z} - \bm{\nabla} h(\bx)~.
\end{equation}
Keeping only terms of second order in $H_{\text{lr},2}$ yields:
\begin{equation}
\begin{split}
H_{\text{lr},2} & \simeq H_{\text{lr}, 2}^{(2, 0)}  = -\frac{1}{2} P^{2} \int {\rm d}^{2}x \int {\rm d}^{2}x' \bigg[\\ &  \frac{\frac{1}{2}\left(\bm{\nabla}h(\bx)- \bm{\nabla}h(\bx')\right)^{2} }{\left|\bx - \bx'\right|^{3}}\\
& + \frac{3\left(h(\bx)- h(\bx') - \bm{\nabla} h(\bx) \cdot \left(\bx - \bx'\right)\right)}{\left|\bx - \bx'\right|^{5}} \\
&\times \left(h(\bx)- h(\bx') - \bm{\nabla} h(\bx') \cdot \left(\bx - \bx'\right)\right) \bigg] ~. \\
\end{split}
\end{equation}
We will denote as $H^{(n, m)}$ the term of order $h^{n} \, u^{m}$ in the expansion of the energy. Introducing the Fourier transform:
\begin{equation}
\label{eq:fourier_h}
h(\bx) = \frac{1}{\sqrt{A}} \sum_{\bq} h_{\bq}\,e^{i\bq \cdot \bx}~,
\end{equation}gives:
\begin{equation}
H_{\text{lr},2}^{(2, 0)} =  -\frac{\pi}{3} P^{2} \sum_{\bq} \left|\bq\right|^{3} \left|h_{\bq}\right|^{2}~.
\end{equation}An explicit calculation of the discrete lattice sum, based on the Ewald method (see the Appendix), actually gives:
\begin{equation}
\label{eq:H_lr2_2_fourier}
H_{\text{lr},2}^{(2, 0)} =  P^{2}  \sum_{\bq} \Big[-\frac{\pi}{3}  \left|\bq\right|^{3} + \frac{1}{a^{3}} f_{2}\left(a \bq\right)\Big]\left|h_{\bq}\right|^{2}~,
\end{equation}
where $a$ is the lattice constant and $f_{2}$ is a regular function of the components of $\bq$ for $a \bq \rightarrow 0$, vanishing as $\left(a q\right)^{4}$ for small wavevectors. The function $f_{2}$ depends on the geometrical details of the lattice. It is in general anisotropic, although it must be invariant under the symmetry operations of the crystal. We will assume that the lattice structure is symmetric enough to ensure that the only invariant quadratic and quartic functions of $\bq$ are $q^{2}$ and $\left(q^{2}\right)^{2}$. Also, we will suppose that the lattice has some inversion or reflection symmetry so that all invariant functions of $\bq$ must be even for $\bq \rightarrow -\bq$. These are actually the same requirements which underlie the isotropy of the elastic and bending coefficients in Eq.~\eqref{eq:H_sr}.

We now turn to the expansion for small fluctuations of $H_{\text{lr},1}$. Expanding up to second order yields:
\begin{equation}
\label{eq:H_lr1_2_real}
\begin{split}
& H_{\text{lr}, 1}  \simeq H_{\text{lr}, 1}^{(0, 0)} + H_{\text{lr}, 1}^{(2, 0)} + H_{\text{lr}, 1}^{(0, 1)} + H_{\text{lr}, 1}^{(0, 2)} \\ & =   \frac{1}{2} \sum_{i}\sum_{j \neq i} p_{i}p_{j} \bigg[ \frac{1}{\left|\bx_{i} -\bx_{j}\right|^{3}}  -\frac{3}{2} \frac{\left(h_{i} - h_{j}\right)^{2}}{\left|\bx_{i} - \bx_{j}\right|^{5}} \\
& -\frac{3}{2} \; \frac{2\left(\bu_{i} -\bu_{j}\right)\cdot \left(\bx_{i} - \bx_{j}\right) + \left(\bu_{i} - \bu_{j}\right)^{2} }{\left|\bx_{i} - \bx_{j}\right|^{5}} \\
& + \frac{15}{2} \frac{\left(\left(\bu_{i} - \bu_{j}\right) \cdot \left(\bx_{i} - \bx_{j}\right)\right)^{2}}{\left|\bx_{i} - \bx_{j}\right|^{7}}\bigg]~.\\
\end{split}
\end{equation}
Fourier transformation gives:
\begin{equation}
\label{eq:H_lr1_2_fourier}
\begin{split}
& H_{\text{lr}, 1}  \simeq \frac{1}{2} \frac{b_{3}}{a} P^{2}  A - \frac{3}{2}  \sum_{i} \sum_{j \neq i} p_{i} p_{j}\frac{\left(\bu_{i} -\bu_{j}\right)\cdot \left(\bx_{i} - \bx_{j}\right)}{\left|\bx_{i} -\bx_{j}\right|^{3}}\\
& + P^{2} \sum_{\bq}\bigg[ -\frac{3}{8} \frac{b_{3}}{a} \,q^{2} + \frac{\pi}{3} \left|\bq\right|^{3} + \frac{1}{a^{3}} f_{1}\left(a \bq\right)\bigg] \left|u_{\alpha,\bq}\right|^{2}\\
 & +  P^{2} \sum_{\bq}\bigg[-\frac{3}{8} \frac{b_{3}}{a} \,q^{2} + \frac{\pi}{3} \left|\bq\right|^{3} + \frac{1}{a^{3}} f_{1}\left(a \bq\right)\bigg]\left|h_{\bq}\right|^{2}\\
 & + P^{2} \sum_{\bq} \bigg[ \frac{15}{16} \frac{b_{3}}{a} q^{2} \left(\frac{1}{2}\delta_{\alpha \beta} + \frac{q_{\alpha} q _{\beta}}{q^{2}}\right)\\
 & - \pi \left|\bq\right|^{3} \Big(\frac{1}{3}\delta_{\alpha \beta} + \frac{q_{\alpha} q_{\beta}}{q^{2}}\Big) +  \frac{1}{a^{3}} g_{\alpha \beta}\Big(a \bq\Big) \bigg]u_{\alpha, \bq}u_{\beta, -\bq}~.
\end{split}
\end{equation}
Details of the calculations are presented in the Appendix. In Eq.~\eqref{eq:H_lr1_2_fourier}, $b_{3}$ is a dimensionless coefficient defined as:
\begin{equation}
\label{eq:b3}
b_{3} \equiv \frac{a}{ P^{2}} \frac{1}{A}\sum_{i}\sum_{j \neq i} p_{i} p_{j} \frac{1}{\left|\bx_{i} - \bx_{j}\right|^{3}}~.
\end{equation}
$f_{1}$ and $g_{\alpha \beta}$ are regular functions of the components of the wavevector at $\bq = 0$. $f_{1}$ and $g_{\alpha \beta}$ are both of order $q^{4}$ at small $\bq$. The linear term in the first line of Eq.~\eqref{eq:H_lr1_2_fourier} represents the tendency of the membrane to stretch under the repulsion between dipole moments. In the continuum limit, it can be written as:
\begin{equation}
\label{eq:linear_u}
\begin{split}
&- \frac{3}{2} \sum_{i} \sum_{j \neq i} p_{i} p_{j}\frac{\left(\bu_{i} -\bu_{j}\right)\cdot \left(\bx_{i} - \bx_{j}\right)}{\left|\bx_{i} -\bx_{j}\right|^{3}} \\ & \approx -\frac{3}{4} \frac{b_{3}}{a} P^{2} \int {\rm d}^{2} x \;\partial_{\alpha} u_{\alpha}\left(\bx\right)~. 
\end{split}
\end{equation}
The electrostatic energy can now be obtained by summing Eqs.~\eqref{eq:H_lr2_2_fourier} and~\eqref{eq:H_lr1_2_fourier}. All terms with a regular small momentum dependence can be represented in real space as a local functional of $\partial_{\alpha}h(\bx)$, $\partial_{\beta}u_{\alpha}(\bx)$ and their gradients. The energy can then be cast into the form:
\begin{equation}
\label{eq:H_lr_2}
\begin{split}
H_{\text{lr}}  & = H_{\text{lr}, 1} + H_{\text{lr}, 2} \\
& \simeq \frac{1}{2} \frac{b_{3}}{a} P^{2} A -\frac{3}{4} \frac{b_{3}}{a} P^{2} \int {\rm d}^{2}x\, u_{\alpha \alpha}\left(\bx\right) \\
& +\int{\rm d}^{2}x \bigg[\frac{1}{2}\kappa_{1}\left(\nabla^{2}h\right)^{2} + \frac{\lambda_{1}}{2}\left(\partial_{\alpha} u_{\alpha}\right)^{2} \\
& + \frac{\mu_{1}}{4}\left(\partial_{\alpha}u_{\beta} + \partial_{\beta}u_{\alpha}\right)\left(\partial_{\alpha}u_{\beta} + \partial_{\beta}u_{\alpha}\right)\bigg] \\
& -\pi P^{2} \sum_{\bq} \left|\bq\right|q_{\alpha} q_{\beta}\,u_{\alpha,\bq}u_{\beta, -\bq}~,
\end{split}
\end{equation}
up to local terms containing higher gradients of the displacement fields~\footnote{The term $-\frac{3}{8}\frac{b_{3}}{a} P^{2} \sum_{\bm{q}} q^{2}\left|h_{\bq}\right|^{2}  $
in Eqs.~\eqref{eq:H_lr_2} and~\eqref{eq:H_lr1_2_fourier} is equivalent to Eq.~(16) of Ref.~\cite{miller_stability_1974} in the case $\epsilon_{a} = \epsilon_{b}$ for the hexagonal lattice geometry considered in Ref.~\cite{miller_stability_1974}. This can be seen by taking advantage of the identity
\begin{equation*}
  \sum_{l, m} \frac{m^{2}}{\left(l^{2} + m^{2}-lm\right)^{\frac{5}{2}}} = \frac{2}{3} \sum_{l, m}\frac{1}{\left(l^{2} + m^{2}-lm\right)^{\frac{3}{2}}}~,
  \end{equation*} where the sum runs over all integer pairs with $(l, m) \neq (0, 0)$. This identity follows from the symmetry of the hexagonal lattice.}. $\kappa_{1}$, $\lambda_{1}$ and $\mu_{1}$ are non-universal constants depending on the details of the lattice geometry. Notice that the negative tension terms
\begin{equation*}
-\frac{3}{8} \frac{b_{3}}{a} P^{2} \sum_{\bq}   q^{2} \left(\left|h_{\bq}\right|^{2} + u_{\alpha, \bq} u_{\alpha, -\bq}\right)
\end{equation*}
in Eq.~\eqref{eq:H_lr2_2_fourier} have been absorbed into the term linear in the strain tensor in Eq.~\eqref{eq:H_lr_2}.

While summing together the electrostatic and the elastic energy, we must set, in Eq.~\eqref{eq:H_sr},
\begin{equation}
\label{eq:tau}
\tau = \frac{3}{4} \frac{b_{3}}{a} P^{2}~,
\end{equation}
so that terms linear in displacement fields disappear from the total energy. The result for the total energy is therefore, in the harmonic approximation:
\begin{equation}
\label{eq:H_2}
\begin{split}
H &= H_{\text{sr}} + H_{\text{lr}} \\
& \simeq  \int {\rm d}^{2}x \,\bigg[\frac{1}{2}\kappa' \left(\nabla^{2} h\right)^{2} + \frac{\lambda'}{2}\left(\partial_{\alpha} u_{\alpha}\right)^{2} \\
& + \frac{\mu'}{4}\left(\partial_{\alpha}u_{\beta} + \partial_{\beta}u_{\alpha}\right)\left(\partial_{\alpha}u_{\beta} + \partial_{\beta}u_{\alpha}\right)\bigg]  \\
& - \pi P^{2} \sum_{\bq} \left|\bq\right|q_{\alpha} q_{\beta} u_{\alpha,\bq}u_{\beta, -\bq}~,
\end{split}
\end{equation}
with values of the elastic coefficients and bending rigidity which are shifted by electrostatic interactions. The most striking feature of Eq.~\eqref{eq:H_2} is that the wavevector dependence of the energy of out-of-plane fluctuations remains proportional to $q^{4}$ and is therefore, as soft as in the purely elastic case.
The cancellation from the Hamiltonian of the tension term~\cite{toner_elastic_1989}, proportional to
\begin{equation*}
\sum_{\bq} q^{2}\left|h_{\bq}\right|^{2} = \int {\rm d}^{2}x \left(\bm{\nabla} h\left(\bx\right)\right)^{2}~,
\end{equation*}
is a consequence of rotational invariance in the ambient space. However,  a term proportional to
\begin{equation}
\label{eq:cubic_term}
\mathcal{O}_{h} = \sum_{\bq} \left|\bq\right|^{3} \left|h_{\bq}\right|^{2}
\end{equation}
is not a priori excluded by rotational symmetry. This is evident from Eq.~\eqref{eq:H_lr_splitted}. Both $H_{\text{lr},1}$ and $H_{\text{lr}, 2}$ are individually rotationally invariant interactions and they both lead to a cubic wavevector dependence of the bending energy. This dependence cancels out only from the sum $H_{\text{lr},1}+H_{\text{lr},2}$. 

The cancellation of the non-local term $\mathcal{O}_{h}$ from the energy of bending undulations could also be predicted by using, instead of Eq.~\eqref{eq:H_lr_splitted}, the equivalent representation of the energy:
\begin{equation}
\label{eq:H_coulomb_repr}
H_{\text{lr}} = \frac{1}{2} \int \frac{{\rm d}^{3}Q }{\left(2\pi\right)^{3}}\;\frac{4\pi}{Q^{2}}\,\rho_{\bm{Q}}\,\rho_{-\bm{Q}}~,
\end{equation}
where
\begin{equation}
\label{eq:rhoq}
\rho_{\bQ} = i \sum_{j} \,p_{j} \bn_{j} \cdot \bQ \;e^{-i \bQ \cdot \br_{j}}~,
\end{equation}
is the Fourier transform in three-dimensional space of the charge density. This method is similar to the approach in Ref.~\cite{peliti_fluctuations_1989}. There, however, approximations were made which lead to the prediction of a $\left|\bm{q}\right|^{3}$ dependence of the quadratic part of the bending energy, in contrast with our results. Eq.~\eqref{eq:H_coulomb_repr} differs from Eq.~\eqref{eq:H_lr_splitted} only because the former includes the infinite energy of self-interaction of dipole moments. This energy, however, is independent on the configuration of the membrane and constitutes a shift of the Hamiltonian by a global constant. Non-local terms in the expansion of Eq.~\eqref{eq:H_coulomb_repr} up to second order in $h$ and $u$ can only arise from the terms:
\begin{equation}
\label{eq:H_lr_2_nonloc}
\begin{split}
&  \frac{1}{2} \int  \frac{{\rm d}^{3}Q}{\left(2 \pi\right)^{3}} \frac{4\pi}{Q^{2}} \left[\rho_{\bQ}^{(1, 0)} \rho_{-\bQ}^{(1, 0)} + \rho_{\bQ}^{(0, 1)} \rho_{-\bQ}^{(0, 1)} \right]  \\ & \equiv \bar{H}_{\text{lr}}^{(2, 0)} + \bar{H}_{\text{lr}}^{(0, 2)}~,
 \end{split}
\end{equation}
where $\rho_{\bQ}^{(1, 0)}$ and $\rho_{\bQ}^{(0, 1)}$ are the terms linear in $h$ and $u$ in the expansion of $\rho_{\bQ}$, respectively. Denoting the in-plane and the out-of-plane components of $\bQ$ as $\bq$ and $q_{z}$, we find for $\bq$ well within the first Brillouin zone:
\begin{equation}
\label{eq:rho_Q_1}
\begin{split}
& \rho_{\bQ}^{(1, 0)} = \sqrt{A}\,P \;Q^{2}h_{\bq}~, \\
& \rho_{\bQ}^{(0, 1)} = \sqrt{A}\,P \;q_{z} \,\bq\cdot \bu_{\bq}~.\\
\end{split}
\end{equation}
Plugging Eq.~\eqref{eq:rho_Q_1} into Eq.~\eqref{eq:H_lr_2_nonloc}, yields:
\begin{equation}
\label{eq:H_lr_2_nonlocal_b}
\begin{split}
 \bar{H}_{\text{lr}}^{(2, 0)} + \bar{H}_{\text{lr}}^{(0, 2)}  & =  \frac{1}{2}   P^{2} \sum_{\bq }\int   \frac{{\rm d}q_{z}}{2 \pi}  \,\frac{4 \pi}{Q^{2}}\Big[ Q^{4} \left|h_{\bq}\right|^{2} \\ & +\left(Q^{2} - q^{2}\right)\left(\bq \cdot \bu_{\bq}\right) \;\left(\bq \cdot \bu_{-\bq}\right)
  \Big]~.
\end{split}
\end{equation}
The fluctuation energy can then be obtained by performing the integral over $q_{z}$ explicitly. In the part quadratic in $h$, $\bar{H}^{(2, 0)}$, the Coulomb interaction factor $1/Q^{2}$ cancels against the factor $Q^{4}$. The resulting interaction is local. The divergence of the $q_{z}$ integral in Eq.~\eqref{eq:H_lr_2_nonlocal_b} is not problematic, because in order to get the full expression for the energy at order $h^{2}$ and $u^{2}$ we still need to add the purely local terms:
\begin{equation*}
\frac{1}{2} \int \frac{ {\rm d}^{3} Q}{\left(2 \pi\right)^{3}} \frac{4\pi}{Q^{2}} \left[2 \rho^{(2, 0)}_{\bQ} \rho^{(0, 0)}_{-\bQ} + 2\rho^{(0, 2)}_{\bQ} \rho^{(0, 2)}_{-\bQ}\right]~,
\end{equation*}
where $\rho_{\bQ}^{(0, 0)}$, $\rho_{\bQ}^{(2, 0)}$ and $\rho_{\bQ}^{(0, 2)}$ are the terms of zeroth and second order in the expansion of $\rho_{\bQ}$. 
Since the energy is finite order by order in the expansion in powers of $u$ and $h$, divergences in the $q_{z}$ integral must cancel out. The only non-local term in Eq.~\eqref{eq:H_lr_2_nonlocal_b}, for which the Coulomb interaction factor $1/Q^{2}$ does not cancel out, is
\begin{equation}
\label{eq:u_propagator_nonloc}
\begin{split}
\bar{\mathcal{O}}_{u} & =  -\frac{1}{2} P^{2} \sum_{\bq} \int \frac{{\rm d}q_{z}}{\left(2 \pi\right)} \,\frac{4 \pi}{Q^{2}} \,q^{2}\left(\bq \cdot \bu_{\bq}\right)\left(\bq \cdot \bu_{-\bq}\right) \\ & = -\pi  P^{2} \sum_{\bq} \,\left|\bq\right|\,q_{\alpha} q_{\beta} \,u_{\alpha, \bq} u_{\beta, -\bq}~. 
\end{split}
\end{equation}
Notice that $\bar{\mathcal{O}}_{u}$ coincides with the non-local term in Eq.~\eqref{eq:H_2}.

The physical interpretation of the cancellation of non-local terms in the quadratic part of the $h$-field energy is more transparent if the previous argument is translated from Fourier to real space. Consider a localized out-of-plane deformation $h\left(\bx\right)$, which is nonzero only within a small region of the membrane. The electrostatic potential at a point $\bm{R}$ in three-dimensional space far from the deformed region is given, in a continuum approximation, by:
\begin{equation}
V\left(\bm{R}\right) = P\int {\rm d}^{2}x \, \frac{\bn\left(\bx\right) \cdot  \left(\bm{R} - \br\left(\bx\right)\right)}{\left|\bm{R}-\br\left(\bx\right)\right|^{3}}~.
\end{equation}
The change in electrostatic potential due to the fluctuation to first order in $h$ can be calculated by making use of the expansion of the normal in Eq.~\eqref{eq:normal_1o}. The result is:
\begin{equation}
\delta V^{(1, 0)}\left(\bm{R}\right) = - P\int {\rm d}^{2}x\frac{\partial}{\partial x_{\alpha}} \left[\frac{h(\bx) \left(X_{\alpha}- x_{\alpha}\right)}{\left|\bm{R} - \bx\right|^{3}}\right]~,
\end{equation}
where $X_{\alpha}$ are the in-plane components of $\bm{R}$. Since the integrand is a total divergence and the deformation $h\left(\bx\right)$ is assumed to be localized, $\delta V^{(1, 0)}$ vanishes. If we now consider a second localized wavepacket, far from the first, we can conclude that there will not be a long-range interaction between the two, at least at the lowest order in the amplitude of the $h$ field. 

 The cancellation of the non-local term $\mathcal{O}_{h}$, Eq.~\eqref{eq:cubic_term}, from the Hamiltonian has crucial physical consequences. If it did not vanish, it would completely modify the critical behaviour, as predicted in Ref.~\cite{toner_elastic_1989}.
 
\subsection{Electrostatic energy: expansion to higher orders}
\label{sec:anharmonic_terms}
We now discuss the properties of higher-order terms in the expansion of the electrostatic energy. For long-wavelength, slowly-varying fluctuations we can calculate the energy approximately by replacing, in Eq.~\eqref{eq:H_lr_splitted}:
\begin{equation}
\label{eq:first_moment_approximation}
\begin{split}
&\br(\bx_{j}) \simeq \br(\bx_{i}) + \partial_{\alpha} \br (\bx_{i}) \left(x_{\alpha i}- x_{\alpha j}\right)~,\\
& \bn_{j} \simeq \bn_{i}~.
\end{split}
\end{equation}
This replacement gives the first order in an expansion in the number of gradients, which translates, in Fourier space, in an expansion in powers of the momentum. The derivative $\partial_{\alpha} \bn$ can be neglected, because it is of higher order. Because the interaction is long-ranged, the gradient series is divergent and presents infinite expansion coefficients. The first order, however, is finite. It gives, therefore, the leading behaviour at long wavelengths of interaction vertices.

When Eq.~\eqref{eq:first_moment_approximation} is substituted into Eq.~\eqref{eq:H_lr_splitted}, we see immediately that $H_{\text{lr}, 2}$ vanishes in the first order of the gradient expansion. This follows from the fact that $\partial_{\alpha} \br(\bx_{i}) \cdot \bn_{i}= 0$. Therefore:
\begin{equation}
\label{eq:anharmonic_terms_local}
\begin{split}
H_{\text{lr}} & \simeq H_{\text{lr}, 1} \simeq \frac{1}{2} \sum_{i, j \neq i} p_{i} p_{j} \Big[ \left(\bx_{i} - \bx_{j}\right)^{2}  
 \\ & + 2u_{\alpha \beta}(\bx_{i})(x_{\alpha i} - x_{\alpha j}) (x_{\beta i }- x_{\beta j})   \Big]^{-\frac{3}{2}}~.
\end{split}
\end{equation}
Expanding in powers of the strain tensor and summing over $j$ leads to a finite local functional of $u_{\alpha \beta}$, which represents the leading part at small wavevectors of the electrostatic energy. Therefore, as anticipated in Section~\ref{sec:harmonic_approximation}, at long-wavelengths, dipolar interactions lead to a shift of anharmonic coupling constants of the short-range Hamiltonian. Non-local behaviour appears as a correction, which is suppressed by a power of the wavevector scale. The first terms of the expansion of Eq.~\eqref{eq:anharmonic_terms_local} are:
\begin{equation}
\label{eq:H_local}
\begin{split}
H_{\text{lr}, 1} & \simeq \frac{1}{2} \frac{b_{3}}{a} P^{2} A  + \frac{b_{3}}{a} P^{2} \int {\rm d}^{2}x \bigg[ -\frac{3}{4} u_{\alpha \alpha} \\ &+ \frac{15}{32}\left(u_{\alpha \alpha}\right)^{2} + \frac{15}{16} \left( u_{\alpha \beta}\right)^{2} \bigg]~,
\end{split}
\end{equation}
if the crystal structure is sufficiently symmetric to ensure the isotropy of  elastic coefficients. Eq.~\eqref{eq:H_local} reproduces the results obtained in Sec.~\ref{sec:harmonic_approximation} for the electrostatic tension $\tau$ and for the shifts of the Lam\'e coefficients $\lambda_{1}$ and $\mu_{1}$, as can be seen by comparison with Eqs.~\eqref{eq:H_lr1_2_fourier} and~\eqref{eq:H_lr_2}.

We now discuss the leading non-local corrections up to order $h^{2} u$ and $h^{4}$. We show that the correction $\bar{\mathcal{O}}_{u}$ in the harmonic part of the energy is promoted to:
\begin{equation}
\label{eq:correction_lame}
\mathcal{O}_{u} = -\pi P^{2} \sum_{\bq} \left|\bq\right| \tilde{u}_{\alpha \alpha}\left(\bq\right) \tilde{u}_{\beta \beta}\left(-\bq\right)~,
\end{equation}
a rotationally invariant interaction, which can be interpreted as a $q$-dependent correction to the Lam\'e coefficient $\lambda$. Up to order $u h^{2}$ and $h^{4}$, $\mathcal{O}_{u}$ is the only non-local interaction to be generated.

In order to calculate the non-local part of interaction vertices, it is convenient to expand the electrostatic energy starting from the expression in Eq.~\eqref{eq:H_coulomb_repr}. Using the expansion for the normal to second order in $h$ and zeroth order in $u$:
\begin{equation}
\begin{split}
\bn\left(\bx \right) & \simeq  \bm{e}_{z} \left(1 - \frac{1}{2}\left(\bm{\nabla} h\left(\bx\right)\right)^{2}\right) - \bm{\nabla} h\left(\bx\right)~,
\end{split}
\end{equation}
we can calculate $\rho_{\bQ}^{(2, 0)}$, the term of order $h^{2}$ in the expansion of the Fourier transform of the charge density $\rho_{\bQ}$:
\begin{equation}
\label{eq:rhoq_2order}
\rho_{\bQ}^{(2, 0)} = -\frac{i}{2} P\,q_{z}\,\sum_{\bk}\left[\bk \cdot \left(\bk + \bq\right) + Q^{2}\right] h_{\bk + \bq} h_{-\bk}~.
\end{equation}
As in Eq.~\eqref{eq:rho_Q_1}, $\bq$ and $q_{z}$ denote the in-plane and the out-of plane components of the three-dimensional wavevector $\bQ$. Again Eq.~\eqref{eq:rhoq_2order} holds under the assumption that $\bq$ lies well within the first Brillouin zone of the lattice. Let us now consider the expansion of the energy, Eq.~\eqref{eq:H_coulomb_repr}, at order $u h^{2}$:
\begin{equation}
\label{eq:Hlr_21}
\begin{split}
H_{\text{lr}}^{(2, 1)}  & = \frac{1}{2} \int   \frac{{\rm d}^{3}Q}{\left(2 \pi\right)^{3}} \frac{8\pi}{Q^{2}}  \Big[\rho_{\bQ}^{(2, 0)}\rho_{-\bQ}^{(0, 1)} \\ &+ \rho_{\bQ}^{(1, 0)}\rho_{-\bQ}^{(1, 1)}  + \rho_{\bQ}^{(2, 1)}\rho_{-\bQ}^{(0, 0)}\Big]~.
\end{split}
\end{equation}
The non-local behaviour of the vertex is only generated by the first term:
\begin{equation}
\label{eq:Hlr_21_bar}
\bar{H} _{\text{lr}}^{(2, 1)} = \frac{1}{2} \int   \frac{{\rm d}^{3}Q}{\left(2 \pi\right)^{3}} \frac{8\pi}{Q^{2}}  \,\rho_{\bQ}^{(2, 0)}\rho_{-\bQ}^{(0, 1)}~.
\end{equation}
The third term in Eq.~\eqref{eq:Hlr_21} is clearly local because fields are inserted at only one of the two $\rho_{\bQ}$. In the second term, the Coulomb interaction $1/Q^{2}$ cancels against the factor $Q^{2}$ in $\rho_{\bQ}^{(1, 0)}$, see Eq.~\eqref{eq:rho_Q_1}. The latter cancellation has the same origin of the cancellation of the propagator correction $\mathcal{O}_{h}$. Similarly, it can be seen that the non-local behaviour of the quartic vertex $H^{(4, 0)}$ is completely encoded in the term:
\begin{equation}
\label{eq:Hlr_40_bar}
\bar{H}_{\text{lr}}^{(4, 0)} = \frac{1}{2} \int   \frac{{\rm d}^{3}Q}{\left(2 \pi\right)^{3}} \frac{4\pi}{Q^{2}}  \;\rho_{\bQ}^{(2, 0)}\rho_{-\bQ}^{(2, 0)}~.   
\end{equation}
Eqs.~\eqref{eq:Hlr_21_bar} and~\eqref{eq:Hlr_40_bar} severely restrict the possible form of the long-range, non-analytic part of interaction vertices. The momentum transfer $\bq$, the only on which the energy can have a singular dependence, can not be exchanged by a single $h$ field, but only by pairs. In real space this can again be visualized by recalling that the potential generated by a localized wavepacket of out-of-plane deformation vanishes to first order in its amplitude.

Combining Eqs.~\eqref{eq:rho_Q_1},~\eqref{eq:rhoq_2order},~\eqref{eq:Hlr_21_bar} and~\eqref{eq:Hlr_40_bar} and keeping only the non-local part of vertices, the part in which the factor $1/Q^{2}$ does not cancel out from the calculations, we obtain:
\begin{equation}
\label{eq:proof_Ou}
\begin{split}
H_{\text{lr}} & \approx  \frac{1}{2} \int \frac{{\rm d}^{3}Q}{\left(2 \pi\right)^{3}} \frac{4 \pi}{Q^{2}} \left(\rho_{\bQ}^{(2, 0)} + \rho_{\bQ}^{(0, 1)}\right) \left(\rho_{-\bQ}^{(2, 0)} + \rho_{-\bQ}^{(0, 1)}\right) \\ & \approx 
- \frac{1}{2} P^{2} \sum_{\bq} \int \frac{{\rm d}q_{z}}{2 \pi} \; \frac{4\pi}{Q^{2}}\; q^{2}\,\tilde{u}_{\alpha \alpha}\left(\bq\right)\,\tilde{u}_{\beta \beta}\left(-\bq\right) \\ & = -\pi  P^{2} \sum_{\bq} \left|\bq\right| \tilde{u}_{\alpha \alpha}\left(\bq\right) \tilde{u}_{\beta \beta}\left(-\bq\right)  = \mathcal{O}_{u}~.
\end{split}
\end{equation}
In Eq.~\eqref{eq:proof_Ou} all equivalencies are intended up to local functionals of the fields. We can now obtain the complete expression for the long-wavelength Hamiltonian by combining Eqs.~\eqref{eq:H_local},~\eqref{eq:proof_Ou} and Eqs.~\eqref{eq:H_sr},~\eqref{eq:tau}. The result is:
\begin{equation}
\label{eq:H_final}
H = \int {\rm d}^{2}x \left[\frac{1}{2}\kappa'\left(\nabla^{2} h\right)^{2} + \frac{\lambda'}{2} \tilde{u}_{\alpha \alpha}^{2} + \mu' \tilde{u}_{\alpha \beta}^{2} \right] + \, \mathcal{O}_{u}~.
\end{equation}
This result extends Eq.~\eqref{eq:H_2}, which was found within the harmonic approximation. 

The non-local term $\mathcal{O}_{u}$ is suppressed by the factor $\left|\bm{q}\right|$ in the long-wavelength limit. It constitutes, therefore, an irrelevant perturbation by power counting in a neighbourhood of $D = 4$, which is the upper critical dimension for crystalline membranes~\cite{aronovitz_fluctuations_1988}. Non-local terms of higher order in powers of the fields $u$ and $h$ have a structure which is similar to $\mathcal{O}_{u}$. They are suppressed by a power $\left|\bm{q}\right|$ with respect to local couplings and they also constitute irrelevant operators by power counting.

The fact that $\mathcal{O}_{u}$ is irrelevant even for physical 2D membranes is suggested by the following considerations. In a scaling analysis, we expect that $\mathcal{O}_{u}$ is irrelevant if $\left|\lambda_{\rm R}(q)\right|, \left|\mu_{\rm R}(q)\right| \gg 2 \pi P^{2} \left|\bm{q}\right|$ at small $q$. This comparison relies on the assumption that $\mathcal{O}_{u}$ is not renormalized, which is suggested by its non-locality. For $D$-dimensional membranes embedded in a $d$-dimensional space, the renormalized elatic moduli $\lambda_{\rm R}\left(q\right)$  and $\mu_{\rm R}\left(q\right)$ vanish as $q^{4-D- 2\eta}$ for $q\rightarrow 0$~\cite{nelson_statistical_1989, aronovitz_fluctuations_1988, bowick_statistical_2001}. The condition $ \left|\lambda_{\rm R}\left(q\right)\right|, \left|\mu_{\rm R}\left(q\right)\right| \gg 2 \pi P^{2}\left|\bm{q}\right| $, then, is fulfilled provided that $\eta > (3 - D)/2$. Field-theoretical calculations~\cite{le_doussal_self-consistent_1992, gazit_structure_2009, le_doussal_anomalous_2018, kownacki_crumpling_2009, guitter_thermodynamical_1989} indicate that $\eta >1/2$ in the case of clean two-dimensional membranes, thus giving an indication that $\mathcal{O}_{u}$ is irrelevant in the physical case $D = 2$, $d = 3$. However, scaling with an exponent $\eta < 1/2$ was predicted to occur in 2D disordered membranes~\cite{le_doussal_anomalous_2018, gornyi_rippling_2015}. In this case our considerations open a question about the possible role of non-local terms.

If, on the other hand, the statistical mechanical properties of the membrane are addressed within the self-consistent screening approximation (SCSA)~\cite{le_doussal_self-consistent_1992, le_doussal_anomalous_2018}, $\mathcal{O}_{u}$ appears to be irrelevant irrespective of the value of $\eta$.  This follows immediately from the fact that $\mathcal{O}_{u}$ is equivalent to a $\bm{q}$-dependent shift of the first Lam\'e coefficient. Within the SCSA, the local Lam\'e coefficient $\lambda'$ and its non-local correction $-2\pi P^{2}\left|\bq\right|$ enter the self-consistent equations only through their sum. For sufficiently small $\bq$, the non-local correction is negligible. The SCSA equations in the long-wavelength limit~\cite{le_doussal_self-consistent_1992, le_doussal_anomalous_2018}, therefore, remain unchanged and the exponent $\eta$ remains necessarily as in the pure short-range case.

\subsection{Shift of elastic coefficients}
We now discuss the shift of elastic coefficients induced by electrostatic interactions. The relations between the shifted elastic constants $\lambda'$ and $\mu'$ and the Lam\'e coefficients $\lambda$ and $\mu$ which appear in the short range Hamiltonian $H_{\text{sr}}$, Eq.~\eqref{eq:H_sr_complete} can be read from Eq.~\eqref{eq:H_lr1_2_fourier}, or equivalently from Eq.~\eqref{eq:H_local}:
\begin{equation}
\label{eq:shift_lame}
\begin{split}
&\lambda' = \lambda + \frac{15}{16} \frac{b_{3}}{a} P^{2}~,\\
&\mu' = \mu + \frac{15}{16}\frac{b_{3}}{a} P^{2}~.
\end{split}
\end{equation}
The shift in the bending rigidity $\kappa' - \kappa$ could be determined from Eq.~\eqref{eq:H_lr2_2_fourier} and ~\eqref{eq:H_lr1_2_fourier} by expanding $f_{2}\left(a \bq \right)$ and $f_{1}\left(a \bq \right)$ up to fourth order in $\bq$. This would require an explicit calculation of $q$-dependent lattice sum, which could be performed with the Ewald method (see the Appendix) with an explicit choice of the Ewald cutoff function $\varphi\left(x\right)$~\footnote{For widely used Ewald cutoff functions, see for example~\cite{aharony_critical_1973, grechnev_thermodynamics_2005, cohen_dipolar_1955}}. In this work, we did not address the renormalization of the bending rigidity explicitly. 

The renormalization of in-plane elastic coefficients, instead, is determined by the simple expression in Eq.~\eqref{eq:shift_lame}. We notice, however, that the 'bare' coefficients $\kappa$, $\lambda$ and $\mu$ do not coincide exactly with the elastic coefficients $\kappa_{0}$, $\lambda_{0}$ and $\mu_{0}$ of the membrane when dipole interactions are switched off. The reason is that dipole-dipole repulsion induces a finite stretching of the membrane which results in an increase of the elastic coefficients. The elastic constants $\kappa$, $\lambda$ and $\mu$, being defined through Eq.~\eqref{eq:H_sr_complete}, represent elastic coefficients after stretching has already occurred.

If the effects of dipole repulsion are weak and the stretching is small so that linear elasticity theory applies while electrostatic interactions are turned on, we can relate the two sets of coefficients through: 
\begin{equation}
\begin{split}
&\kappa = \xi^{2} \kappa_{0}~, \\
&\lambda = \xi^{2} \lambda_{0}~, \\
& \mu = \xi^{2} \mu_{0}~,
\end{split}
\end{equation}
where $\xi$ is the stretching factor, determined by:
\begin{equation}
\label{eq:stretching_factor}
\tau = \frac{3}{4} \frac{b_{3}}{a} P^{2} = \left(\lambda_{0} + \mu_{0}\right) \xi^{2}\left(\xi^{2} - 1\right)~.
\end{equation}
Here $a$ and $P$ are the 'final' values of the lattice constant and the polarization density, calculated at the equilibrium value of the stretching factor in presence of electrostatic interactions, the same $a$ and $P$ which were used throughout the rest of this paper. By the previous considerations, Eq.~\eqref{eq:stretching_factor} holds under the assumption $\xi -1 \ll 1$.

We now give an order of magnitude estimate of the effects of dipolar interactions on the elastic parameters of the membrane. We consider, as an example, the case of graphene derivatives involving Group IA and Group VIIA atoms~\cite{klintenberg_theoretical_2010}. In these materials alkali and halogen atoms form partially ionic bonds with carbon atoms in the graphene layer. If alkali atoms bind on one side of the graphene layer and halogen atoms on the opposite side, an out-of-plane permanent polarization results. This occurs, for example, in C$_{2}$HF and C$_{2}$LiF in the 'chair' geometry described in Ref.~\cite{klintenberg_theoretical_2010}. For these materials, we expect that the typical scale of the polarization density $P$ is roughly 1 e/\AA. From Eq.~\eqref{eq:shift_lame}, it follows that corrections to the Lam\'e coefficients are of the order of 1 eV/\AA$^{2}$. They are, therefore, of the same order of magnitude of the 'bare' elastic coefficients $\lambda$ and $\mu$. The same considerations suggest that the shift of the bending rigidity is of the order of 1 eV.

More specifically, we expect that among the class of graphene derivatives involving Group IA and Group VIIA atoms, the largest dipole moments and the strongest polarization effects arise in C$_{2}$LiF~\cite{klintenberg_theoretical_2010}. An estimate of the shift of elastic coefficient can be done, within our model, by describing C$_{2}$LiF (in the 'chair' geometry) as a lattice with dipole moments $ q_{\text{Li}} d_{\text{Li}}$ and $-q_{\text{F}} d_{\text{F}}$ located at the two sublattices of the hexagonal graphene structure. For the charge transfer to lithium and fluorine atoms, we assume in an order of magnitude estimate, $q_{\text{Li}} = +\text{e}$ and $q_{\text{F}} = -\text{e}$. The structural properties of C$_{2}$LiF were calculated in Ref.~\cite{klintenberg_theoretical_2010}. The bond lengths $d_{\text{Li}}$ and $d_{\text{F}}$ and the lattice constant $a$ are 2.16 \AA, 1.44 \AA~and 2.59 \AA, respectively. We obtain a polarization density $P \simeq $ 0.6 e/\AA. The value of the lattice sum $b_{3}$ is close to $9$~\cite{grechnev_thermodynamics_2005} in the case of a square lattice and we expect that it remains of the same order in a hexagonal lattice. Eq.~\eqref{eq:shift_lame} then leads to corrections to the Lam\'e coefficients of the order of 10 eV/\AA$^{2}$. This result is probably an overestimation. It suggests, however, that electrostatic effects can play an important role in the elastic properties of C$_{2}$LiF.

 \section{Conclusions}
In conclusion, we addressed the effects of dipole-dipole interactions on the long-wavelength thermal fluctuations of membranes exhibiting a permanent out-of-plane polarization directed at each point along the local normal. We focused on crystalline membranes in the flat phase, a case which is particularly important for the thermodynamic properties of two-dimensional materials at room temperature.

We found that, even in presence of long-range dipole-dipole interactions, the dispersion relation of bending fluctuations remains as soft as in the short-range case and can be described at long wavelengths in terms of a bending rigidity. Furthermore, the leading behaviour of the electrostatic energy in the limit of long wavelengths can be represented through a local functional of the strain tensor. These results have important consequences in the statistical mechanical properties of the membrane. They imply that the effective Hamiltonian, in the leading order in the limit of long wavelengths, has the same local form of the Hamiltonian of elasticity theory. Non-local terms are suppressed in the long-wavelength limit and they are irrelevant by power counting. The expected result, therefore, is that the large-distance behaviour and the scaling exponents of elastic membranes are not modified by dipole-dipole interactions, at least in a neighbourhood of $D=4$. A scaling analysis and the self-consistent screening approximation support the expectation that the non-local correction of smallest order in powers of the displacement fields remains irrelevant even for $D = 2$, at least in the case of clean membranes, for which the exponent $\eta$ is larger than $1/2$.  The condition $\eta >1/2$, however, can be violated in disordered membranes~\cite{le_doussal_anomalous_2018, gornyi_rippling_2015}. In this case, the role of dipole-dipole interactions in the scaling behaviour raises a new question.

We believe that the model which we studied in this work can help to understand thermal fluctuations of two-dimensional materials, such as the graphene derivatives CF, C$_{2}$HF and C$_{2}$LiF~\cite{klintenberg_theoretical_2010}, which exhibit permanent out-of-plane polarization. More and more 2D materials with finite out-of-plane polarization have been predicted or discovered in recent years, including, for example, MoSSe~\cite{riis-jensen_efficient_2018} and WTe$_{2}$~\cite{fei_ferroelectric_2018}.

In addition, we addressed the shift of elastic coefficients due to long-range interactions. As a model for crystalline membranes with out-of-plane polarization, we considered a deformable elastic sheet with an embedded lattice of point dipole moments. We derived explicit expressions for the shift of the Lam\'e coefficients $\lambda$ and $\mu$ induced by electrostatics and an implicit expression for the renormalization of the bending rigidity. We estimated that the  renormalization of elastic moduli can be important in graphene derivatives such as C$_{2}$LiF~\cite{klintenberg_theoretical_2010}, in which polarization arises from charge transfer between carbon, alkali and halogen atoms which form partially ionic bonds. Analytic expressions for the shift of elastic coefficients in our model could be useful, for example, as a basis for comparison with computational results for the structural properties of 2D materials.

\begin{acknowledgments}
This work was supported by the Netherlands Organisation for Scientific Research (NWO) via the Spinoza Prize.
\end{acknowledgments}

\appendix*
\section{Harmonic part of the electrostatic energy: details of the calculations}
All calculations in Section~\ref{sec:harmonic_approximation} rely on Fourier transforms of power-law interactions on a lattice. It is convenient to begin by studying the sum:
\begin{equation}
\label{eq:S_3}
S_{n}\left(\bq\right) \equiv \frac{a^{n -2}}{P^{2}} \frac{1}{A}\sum_{i}\sum_{j \neq i}p_{i} p_{j} \frac{e^{i \bq \cdot \left(\bx_{i} - \bx_{j}\right)}}{\left|\bx_{i} - \bx_{j}\right|^{n}}
\end{equation}
for $n = 3$. As in the main text, $A$ is the total area, $P$ the polarization density and $a$  is the lattice constant. The interesting, long-range, character of power-law interactions results in a non-analytic behaviour of $S_{n}\left(\bq\right)$ as $\bq \rightarrow 0$. The non-analyticity arises from the large distance part of the sum in Eq.~\ref{eq:S_3}, where the summation can be accurately approximated with a continuum integral. Replacing the lattice sum with an integral is however impossible at small $\left|\bx_{i} - \bx_{j}\right|$, because of the short-distance divergence. It is convenient, therefore to make use of an Ewald method~\cite[See e.g.][]{aharony_critical_1973, grechnev_thermodynamics_2005, cohen_dipolar_1955}. We decompose $S_{3}\left(\bq\right)$ into a short-range part $S_{3}^{<}\left(\bq\right)$ and a long-range part $S_{3}^{>}\left(\bq\right)$:
\begin{equation}
\begin{split}
S_{3}\left(\bq\right) & = S_{3}^{<}\left(\bq\right) + S_{3}^{>}\left(\bq\right)~,\\
S_{3}^{<}\left(\bq\right)  & \equiv \frac{a}{P^{2}} \frac{1}{A}\sum_{i}\sum_{j \neq i}p_{i} p_{j} \frac{e^{i \bq \cdot \left(\bx_{i} - \bx_{j}\right)}}{\left|\bx_{i} - \bx_{j}\right|^{3}} \\ & \times \varphi \left(\frac{\left|\bx_{i} - \bx_{j}\right|}{a}\right)~,\\
S_{3}^{>}\left(\bq\right)  & \equiv \frac{a}{P^{2}} \frac{1}{A}\sum_{i}\sum_{j \neq i}p_{i} p_{j} \frac{e^{i \bq \cdot \left(\bx_{i} - \bx_{j}\right)}}{\left|\bx_{i} - \bx_{j}\right|^{3}} \\
&\times \left[1 - \varphi \left(\frac{\left|\bx_{i} - \bx_{j}\right|}{a}\right)\right]~.
\end{split}
\end{equation}
Here $\varphi$ is a cutoff function, which vanishes faster than any power of $\left|\bx\right|$ for $\left|\bx\right|\rightarrow \infty$ and such that $1-\varphi\left(\left|\bx\right|/a \right)$ vanishes faster than $\left|\bx\right|$ at short distances. The particular choice of $\varphi$ is unimportant in the following discussion. $S_{3}^{<}\left(\bq\right)$, due to the large-distance cutoff, is a regular function of the components of its vector argument. The non-analytic wavevector dependence arises only from $S_{3}^{>}\left(\bq\right)$. In order to calculate $S_{3}^{>}\left(\bq\right)$, it is convenient to transform the sum over the direct lattice into a sum over reciprocal lattice vectors, by making use of the Poisson summation formula. For a generic function, this transformation reads~\cite{aharony_critical_1973, grechnev_thermodynamics_2005, cohen_dipolar_1955}:
\begin{equation}
\label{eq:poisson}
\begin{split}
& \frac{a}{P^{2} A} \sum_{i} \,\sum_{j} p_{i}p_{j} f\left(\bx_{i} - \bx_{j}\right)  \\
 & = \frac{a}{P^{2} \Omega} \sum_{\bx_{\ell}} \,\sum_{i, j\in \Omega_{0}} p_{i} p_{j} f\left(\bx_{\ell} + \bt_{i} - \bt_{j}\right) \\
& = \frac{a}{P^{2} \Omega^{2}} \sum_{\bg_{m}}  \,\sum_{i, j\in \Omega_{0}} p_{i} p_{j}\, e^{i \bg_{m}\cdot\left(\bt_{i}- \bt_{j}\right)} \int {\rm d}^{2} x \,f(\bx) \,e^{-i \bg_{m} \cdot \bx}~.
\end{split}
\end{equation}
Here $\Omega$ is the area of the unit cell of the Bravais lattice, $\bx_{\ell}$ are primitive vectors of the direct Bravais lattice and $\bg_{m}$ are reciprocal lattice vectors. In the second and the third line of Eq.~\eqref{eq:poisson}, the sum over $i$ and $j$ is restricted to the dipoles lying within the first unit cell, denoted as $\Omega_{0}$. The positions of dipoles within the cell $\Omega_{0}$ are the generating vectors $\bt_{i}$. Eq.~\eqref{eq:poisson} can be directly applied to the evaluation of $S_{3}^{>}\left(\bq\right)$ by choosing:
\begin{equation}
f\left(\bx\right) = \frac{e^{i \bq \cdot \bx}}{\left|\bx\right|^{3}}\,\left(1- \varphi\left(\frac{\left|\bx\right|}{a}\right)\right)~.
\end{equation}
The transformation gives:
\begin{equation}
\label{eq:S3_poisson}
\begin{split}
& S_{3}^{>}\left(\bq\right)  = \frac{a}{P^{2}} \frac{1}{\Omega^{2}} \sum_{\bg_{m}} \sum_{i, j\in \Omega_{0}} p_{i} p_{j}\, e^{i \bg_{m}\cdot\left(\bt_{i}- \bt_{j}\right)} \\
&  \times \int {\rm d}^{2} x \;\frac{e^{i \left(\bq -\bg_{m}\right)\cdot \bx}}{\left|\bx\right|^{3}}\left(1- \varphi\left(\frac{\left|\bx\right|}{a}\right)\right)~.
\end{split}
\end{equation}
In Eq.~\eqref{eq:S3_poisson}, all terms with a finite $\bg_{m}$ are regular functions of $\bq$ for $\bq\rightarrow 0$. Singular behaviour can only occur in the $\bg_{m} = 0$ term of the sum~\cite{aharony_critical_1973}. This term coincides precisely with the continuum integral approximation to $S_{3}^{>}\left(\bq\right)$. Therefore:
\begin{equation}
\label{eq:S3_cont}
\begin{split}
S_{3}^{>}(\bq) &= a \int {\rm d}^{2}x \; \frac{e^{i \bq \cdot \bx}}{\left|\bx\right|^{3}}\left(1- \varphi\left(\frac{\left|\bx\right|}{a}\right)\right)  + r\left(a \bq\right)~,
\end{split}
\end{equation}
where $r\left(a \bq \right)$ is a regular function.

In order to calculate the integral in Eq.~\eqref{eq:S3_cont}, it is useful to make use of the Fourier transform:
\begin{equation}
\label{eq:fourier_cutoff}
 \int {\rm d}^{2}x \;\frac{e^{-i\bk\cdot \bx}}{\left(\bx^{2} + \varepsilon^{2}\right)^{\frac{3}{2}}} = \frac{2\pi}{\varepsilon} \, e^{-\varepsilon \left|\bk\right|}~.
\end{equation}
The integral can be written as:
\begin{equation}
\label{eq:uv_regularization_temp}
\begin{split}
    & S_{3}^{>}\left(\bq\right)  - r\left(a \bq\right) =  \\
    & = a  \lim_{\varepsilon \rightarrow 0} \left[\frac{2 \pi}{\varepsilon} e^{-\epsilon \left|\bq\right|} - \int {\rm d}^{2}x \,\frac{e^{i \bq \cdot \bx} \,\varphi(\left|\bx\right|/a)}{\left(\bx^{2} + \varepsilon^{2}\right)^{\frac{3}{2}}}\right]~.\\
\end{split}
\end{equation}
The second term in the left-hand side of Eq.~\eqref{eq:uv_regularization_temp} is a regular function of $\bq$, due to the large-distance cutoff. In addition, the limit $\varepsilon \rightarrow 0$ must be finite, because the initial expression in Eq.~\eqref{eq:S3_cont}, in which $\varepsilon = 0$ is well defined. As a result:
\begin{equation}
S_{3}^{>}\left(\bq\right) = r'\left(a \bq \right) - 2\pi a \left|\bq\right| ~,
\end{equation}
where $r'\left(a \bq \right)$ is a new regular function of the wavevector. Summing $S_{3}^{<}(\bq)$ and $S_{3}^{>}(\bq)$ finally gives:
\begin{equation}
\label{eq:S3_final}
S_{3}\left(\bq\right) = b_{3} - 2 \pi a \left|\bq \right| + s_{3}\left( a \bq \right)~.
\end{equation}
$s_{3}$ is regular and vanishes as $q^{2}$ at small $\bq$. The constant term $b_{3}= S_{3}(0)$ coincides with the coefficient defined in Eq.~\eqref{eq:b3} in the main text.

The calculations in Section~\ref{sec:harmonic_approximation} require, besides $S_{3}\left(\bq\right)$, also the lattice sums $S_{5}\left(\bq\right)$ and $S_{7}\left(\bq\right)$. These are related to $S_{3}(\bq)$ through:
\begin{equation}
\label{eq:recursion_Sn}
    -\frac{1}{a^{2}}\frac{\partial}{\partial q_{\alpha}} \frac{\partial }{\partial q_{\alpha}} S_{n+2}\left(\bq\right) = S_{n}\left(\bq\right)~.
\end{equation}
If the functions $S_{n}\left(\bq\right)$ were perfectly isotropic, Eq.~\eqref{eq:recursion_Sn} would be sufficient to determine the coefficient of $\left|\bq\right|^{m + 2}$ in $S_{n+2}\left(\bq\right)$ from the coefficient of $\left|\bq\right|^{m}$ in $S_{n}\left(\bq\right)$. In general, $S_{n}\left(\bq\right)$ are anisotropic, although they exhibit the same symmetry of the crystal. However, the non-analytic part of any $S_{n}\left(\bq\right)$ must be perfectly isotropic. The steps which led to the calculation of $S_{3}\left(\bq\right)$ could be repeated similarly for arbitrary $n$. The singular dependence on $\bq$ arises only from a term analogue to Eq.~\eqref{eq:S3_cont}, a continuum isotropic integral with no traces of the crystal structure anisotropy. For the non-singular part, we will make the assumption that the symmetry of the crystal is large enough to forbid anisotropic functions at least up to fourth order in $\bq$. Under this assumption, using Eq.~\eqref{eq:recursion_Sn} recursively yields:
\begin{equation}
\label{eq:S5_S7}
\begin{split}
     S_{5}\left(\bq\right) = b_{5} - \frac{1}{4} & b_{3} a^{2} q^{2} + \frac{2}{9} \pi a^{3} \left|\bq\right|^{3} + s_{5}\left(a \bq \right)~,\\
    S_{7}\left(\bq\right) = b_{7} - \frac{1}{4} & b_{5} a^{2} q^{2} + \frac{1}{64} b_{3} a^{4} q^{4} - \frac{2 \pi}{225} a^{5} \left|\bq\right|^{5} \\ & + s_{7}\left(a \bq \right)~,
\end{split}
\end{equation}
where $s_{5}$ and $s_{7}$ are regular and vanish as $q^{4}$ and $q^{6}$ respectively. Now all necessary ingredients for the calculation of $H_{\text{lr}, 1}^{(2)}$ and $H_{\text{lr}, 2}^{(2, 0)}$ have been derived. Fourier transformation of $H_{\text{lr}, 2}^{(2, 0)}$ gives:
\begin{equation}
\begin{split}
H_{\text{lr}, 2}^{(2, 0)}  & = -\frac{1}{2}\sum_{i} \sum_{j \neq i} p_{i} p_{j} \bigg[
\frac{\frac{1}{2}\left(\bm{\nabla}h\left(\bx_{i}\right) - \bm{\nabla}h\left(\bx_{j}\right)\right)^{2}}{\left|\bx_{i} - \bx_{j}\right|^{3}}
\\ & + \frac{3 \left(h_{i} - h_{j} - \bm{\nabla}h(\bx_{i})\cdot \left(\bx_{i} - \bx_{j}\right)\right)}{\left|\bx_{i} - \bx_{j}\right|^{5}}  \\ & \times \left(h_{i} - h_{j} - \bm{\nabla}h(\bx_{j})\cdot \left(\bx_{i} - \bx_{j}\right)\right)\bigg]\\
& H_{\text{lr}, 2}^{(2, 0)}  = \frac{1}{2}\frac{P^{2}}{a} \sum_{\bq} \Big[q^{2}\left(S_{3}(\bq) - S_{3}(0)\right) \\ & + \frac{6}{a^{2}} \left(S_{5}(\bq) - S_{5}(0)\right) - \frac{6}{a^{2}}  q_{\alpha} \frac{\partial}{\partial q_{\alpha}}S_{5}(\bq) \\ & + \frac{3}{a^{2}} q_{\alpha} q_{\beta} \frac{\partial}{\partial q_{\alpha}} \frac{\partial}{ \partial q_{\beta}} S_{5}(\bq)\Big] \left|h_{\bq}\right|^{2} \\
& = P^{2} \sum_{\bq} \Big[- \frac{\pi}{3} \left|\bq\right|^{3} + \frac{1}{a^{3}} f_{2}\left(a \bq \right)\Big]\left|h_{\bq}\right|^{2}~,
\end{split}
\end{equation}
where $f_{2}(a \bq)$ is an analytic function at $\bq = 0$, which vanishes as $q^{4}$ for $\bq \rightarrow 0$. The Fourier transform of $H_{\rm lr, 1}$ is calculated similarly. From Eq.~\eqref{eq:H_lr1_2_real} in the main text:

\begin{equation}
\begin{split}
& H_{\text{lr}, 1}  \simeq H_{\text{lr}, 1}^{(0, 0)} + H_{\text{lr}, 1}^{(2, 0)} + H_{\text{lr}, 1}^{(0, 1)} + H_{\text{lr}, 1}^{(0, 2)} \\
&  = \frac{1}{2}\frac{b_{3}}{a} P^{2} A - \frac{3}{2}\sum_{i} \sum_{j \neq i} p_{i} p_{j}\frac{\left(\bu_{i} -\bu_{j}\right)\cdot \left(\bx_{i} - \bx_{j}\right)}{\left|\bx_{i} -\bx_{j}\right|^{3}}\\
 & +\frac{P^{2}}{2 a}  \sum_{\bq} \bigg[ \frac{3}{a^{2}} \left(S_{5}(\bq) - S_{5}(0) \right) \left|h_{\bq}\right|^{2} \\
  & + \frac{3}{a^{2}}  \left(S_{5}(\bq) - S_{5}(0) \right) u_{\alpha, \bq}u_{\alpha, -\bq} \\
  & + \frac{15}{a^{4}} \left(\frac{\partial^{2} S_{7}(\bq)}{\partial q_{\alpha} \partial q_{\beta}} - \frac{\partial^{2} S_{7}(0)}{\partial q_{\alpha} \partial q_{\beta}} \right)u_{\alpha, \bq} u_{\beta, -\bq}\bigg]~.
\end{split}
\end{equation}
Using Eq.~\eqref{eq:S3_final} and Eq.~\eqref{eq:S5_S7} leads to the result in the main text, Eq.~\eqref{eq:H_lr1_2_fourier}.

\end{document}